# Determination of Complex Refractive Index using Prism composed of Absorbing Medium

Shosuke Sasaki

Center for Advanced High Magnetic Field Science, Graduate School of Science, Osaka University, Toyonaka, Osaka 560-0043, Japan

E-mail: sasaki@mag.ahmf.sci.osaka-u.ac.jp

Abstract

In light refraction between two transparent media, Snell's law describes the relationship between the incident angle and refraction angle. The refractive index is usually determined from Snell's law using the minimum deviation angle through a prism. A medium that partially absorbs light has complex permittivity (or permeability), and its refractive index has an imaginary part. Because of the imaginary part, Snell's law does not hold for light-absorbing media. Accordingly, other methods is required to determine the complex value of the refractive index in a light-absorbing medium. The exact solution of Maxwell's equations was obtained for light passing through a prism composed of a light-absorbing medium. Thereby the optical path was determined. The difference of optical paths between light-absorbing medium and transparent medium was calculated numenically. The difference is very large in the neighborhood of perfect reflection area of transparent medium. We found two methods to determine the complex value of the refractive index using the optical path. Both methods analyzed the solutions in detail for two specific incident angles. The deviation angles corresponding to the two incident angles determined the complex value of the prism's refractive index. The obtained refractive index is independent of the apex angle of the prism.

## 1. Introduction

In previous papers [1] and [2], the electromagnetic field was examined for a system with a boundary between transparent medium A and light-absorbing medium B. The present study clarifies the light path through a prism made of a substance that partially absorbs light. Maxwell's equations were solved for the system. The obtained light path is different from Snell's law because of the ligh absorption. Moreover, the exit wave was not a typical plane wave. The electromagnetic wave exiting the prism has a wave vector with an imaginary part, even in vacuum, because of the ligh absorption.

The reflection and refraction of light in transparent media have been rigorously explained using Maxwell's equations [3, 4, 5]. The refractive index is usually determined by employing the minimum value $\delta_0$ of the deviation angle $\delta$ through a prism with apex angle $\alpha$, as shown in Fig.1. The refractive index $n$ is given by;

$$n = \frac{\sin((\delta_0+\alpha)/2)}{\sin(\alpha/2)} \qquad (1)$$

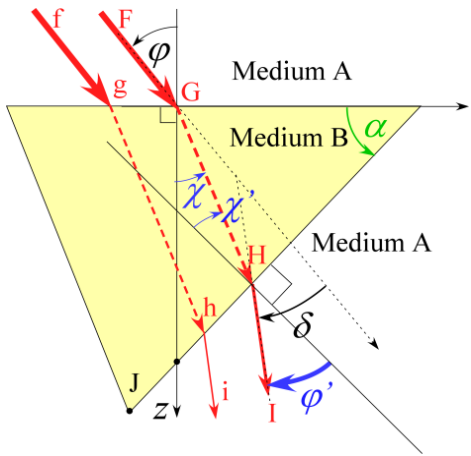

Fig.1 Light paths through a prism: $\varphi$ is an incident angle. $\delta$ is the deviation angle. $\alpha$ represents the apex angle of the prism.

The obtained value $n$ is independent of the apex angle $\alpha$ and depends only on the prism's material. Because the right-hand side of Eq.(1) is a real value, we cannot apply it to substances with an imaginary part of the permittivity and/or magnetic permeability. In other words, the traditional method cannot be applied to light-absorbing media.

We propose a new method to determine both the real and imaginary parts of the refractive index by measuring two specific deviation angles using a prism made of a light-absorbing medium. The complex value obtained was independent of the apex angle of the prism.

## 2. Deviation Angle of Exit Wave through Prism

Figure 1 shows a schematic diagram of light passing through a prism with apex angle $\alpha$. The *x*-axis is to the right, the *y*-axis is from the back to the front of the paper, and the *z*-axis is at the bottom. Let $t$, $\boldsymbol{r}$ and $\varphi$ be the time, position vector, and incident angle to the surface of the prism, respectively. Also, $\omega$ and $\boldsymbol{k}$ represent the angular frequency and wave vector of the incident plane wave. The incident electromagnetic fields $\boldsymbol{E}$ and $\boldsymbol{B}$ are expressed as follows [1], [2]:

$$\boldsymbol{E} = \widehat{\boldsymbol{E}} e^{i(\boldsymbol{k}\cdot\boldsymbol{r}-\omega t)}, \quad \boldsymbol{B} = \frac{1}{\omega}\boldsymbol{k}\times\boldsymbol{E} \quad \text{for } z<0 \tag{2a}$$

$$\boldsymbol{k} = \frac{\omega}{c_0}\sqrt{\hat{\varepsilon}_A\,\hat{\mu}_A}(\sin\varphi, 0, \cos\varphi), \quad c_0 = \frac{1}{\sqrt{\varepsilon_0\mu_0}} \tag{2b}$$

where $\varepsilon_0, \mu_0$ and $c_0$ are the permittivity, magnetic permeability, and light velocity in a vacuum, respectively [3, 4, 5]. $\widehat{\boldsymbol{E}}$ is the constant coefficient vector of the electric field, which is orthogonal to the wave vector $\boldsymbol{k}$. Additionally, $\hat{\varepsilon}_A$ and $\hat{\mu}_A$ are the dielectric constant and relative permeability of medium A, respectively.

Next, the electromagnetic fields $\boldsymbol{E}''$ and $\boldsymbol{B}''$ of the refraction wave are expressed by employing the wave vector $\boldsymbol{K}$ and angular frequency $\omega$ as follows:

$$\boldsymbol{E}'' = \widehat{\boldsymbol{E}}'' e^{i(\boldsymbol{K}\cdot\boldsymbol{r}-\omega t)}, \quad \boldsymbol{B}'' = \frac{1}{\omega}\boldsymbol{K}\times\boldsymbol{E}'' \quad \text{for } z>0 \tag{3}$$

In the previous papers [1] and [2], we separately examined dielectric and magnetic materials. Here, we examine a more general case with complex values for both the dielectric constant and the relative permeability. Maxwell's equations derive the following wave equation for the electric field, $\boldsymbol{E}''$:

$$\Delta \boldsymbol{E}'' - \varepsilon_0 \hat{\varepsilon}_B \mu_0 \hat{\mu}_B \frac{\partial^2 \boldsymbol{E}''}{\partial t^2} = 0 \tag{4}$$

Here, $\hat{\varepsilon}_B$ and $\hat{\mu}_B$ are the dielectric constant and relative permeability of medium B which have real and imaginary parts, respectively, as

$$\hat{\varepsilon}_B = \hat{\varepsilon}_B^{Re} + i\hat{\varepsilon}_B^{Im}, \quad \hat{\mu}_B = \hat{\mu}_B^{Re} + i\hat{\mu}_B^{Im} \tag{5a}$$

where the four values $\hat{\varepsilon}_B^{Re}$, $\hat{\varepsilon}_B^{Im}$, $\hat{\mu}_B^{Re}$ and $\hat{\mu}_B^{Im}$ are real numbers [6], [7]. Substitution of Eq.(3) into Eq.(4) yields

$$\boldsymbol{K} \cdot \boldsymbol{K} = \varepsilon_0 \mu_0 \omega^2 \left( \hat{\varepsilon}_B^{Re} \hat{\mu}_B^{Re} - \hat{\varepsilon}_B^{Im} \hat{\mu}_B^{Im} + i\left( \hat{\varepsilon}_B^{Im} \hat{\mu}_B^{Re} + \hat{\varepsilon}_B^{Re} \hat{\mu}_B^{Im} \right) \right) \tag{5b}$$

The *x*- and *y*-components of the electric field are continuous at the boundary $z = 0$:

$$K_x = k_x = \frac{\omega}{c_0} \sqrt{\hat{\varepsilon}_A \hat{\mu}_A} \sin\varphi, \tag{6a}$$

$$K_y = k_y = 0 \tag{6b}$$

Substitution of Eqs.(6a) and (6b) into the left-hand side of Eq.(5b) yields

$$\begin{aligned}K_z^2 &= (K_z^{Re} + iK_z^{Im})^2 \\ &= \varepsilon_0 \mu_0 \omega^2 \left( \hat{\varepsilon}_B^{Re} \hat{\mu}_B^{Re} - \hat{\varepsilon}_B^{Im} \hat{\mu}_B^{Im} - \hat{\varepsilon}_A \hat{\mu}_A \sin^2\varphi + i\left( \hat{\varepsilon}_B^{Im} \hat{\mu}_B^{Re} + \hat{\varepsilon}_B^{Re} \hat{\mu}_B^{Im} \right) \right)\end{aligned} \tag{7}$$

We introduce an auxiliary variable $\zeta$ as

$$\zeta = \arctan\{ \left( \hat{\varepsilon}_B^{Im} \hat{\mu}_B^{Re} + \hat{\varepsilon}_B^{Re} \hat{\mu}_B^{Im} \right) / \left( \hat{\varepsilon}_B^{Re} \hat{\mu}_B^{Re} - \hat{\varepsilon}_B^{Im} \hat{\mu}_B^{Im} - \hat{\varepsilon}_A \hat{\mu}_A \sin^2\varphi \right) \} \tag{8}$$

Then the wave vector $\boldsymbol{K}$ is obtained as

$$\boldsymbol{K} = \frac{\omega}{c_0} \left( \sqrt{\hat{\varepsilon}_A \hat{\mu}_A} \sin\varphi, 0, \left\{ \left( \hat{\varepsilon}_B^{Re} \hat{\mu}_B^{Re} - \hat{\varepsilon}_B^{Im} \hat{\mu}_B^{Im} - \hat{\varepsilon}_A \hat{\mu}_A \sin^2\varphi \right)^2 + \left( \hat{\varepsilon}_B^{Im} \hat{\mu}_B^{Re} + \hat{\varepsilon}_B^{Re} \hat{\mu}_B^{Im} \right)^2 \right\}^{1/4} e^{i\zeta/2} \right) \tag{9}$$

The imaginary part of $\boldsymbol{K}$ appears only in the *z*-component; therefore, attenuation occurs only in the *z*-direction. This result is in accordance with the previous papers [1] and [2] in light-absorbing medium B. The real part of $\boldsymbol{K}$ produces a phase change; therefore, the refraction angle $\chi$ in Fig.1 is derived from

$$\tan\chi = \frac{K_x}{K_z^{\text{Re}}} = \frac{\sqrt{\hat{\varepsilon}_A \hat{\mu}_A}\sin\varphi}{\left\{\left(\hat{\varepsilon}_B^{\text{Re}}\hat{\mu}_B^{\text{Re}} - \hat{\varepsilon}_B^{\text{Im}}\hat{\mu}_B^{\text{Im}} - \hat{\varepsilon}_A\hat{\mu}_A\sin^2\varphi\right)^2 + \left(\hat{\varepsilon}_B^{\text{Im}}\hat{\mu}_B^{\text{Re}} + \hat{\varepsilon}_B^{\text{Re}}\hat{\mu}_B^{\text{Im}}\right)^2\right\}^{1/4}\cos(\zeta/2)} \quad (10a)$$

Accordingly the refraction angle $\chi$ is given by

$$\chi = \arctan\left[\frac{\sqrt{\hat{\varepsilon}_A \hat{\mu}_A}\sin\varphi}{\left\{\left(\hat{\varepsilon}_B^{\text{Re}}\hat{\mu}_B^{\text{Re}} - \hat{\varepsilon}_B^{\text{Im}}\hat{\mu}_B^{\text{Im}} - \hat{\varepsilon}_A\hat{\mu}_A\sin^2\varphi\right)^2 + \left(\hat{\varepsilon}_B^{\text{Im}}\hat{\mu}_B^{\text{Re}} + \hat{\varepsilon}_B^{\text{Re}}\hat{\mu}_B^{\text{Im}}\right)^2\right\}^{1/4}\cos(\zeta/2)}\right] \quad (10b)$$

(*Short summary 1*): The refraction angle $\chi$ in Fig.1 was determined for any value of incident angle $\varphi$ using $\hat{\varepsilon}_A$, $\hat{\varepsilon}_B$, $\hat{\mu}_A$, $\hat{\mu}_B$ of media A and B.

Next, we examined the electromagnetic fields in the outer region where light exited from the prism. The region was filled with medium A which is the same substance in the incident region. However, Eqs.(2a) and (2b) cannot be applied, because the light intensity changes along a direction perpendicular to the travelling direction of the light. We studied the exact solution for the electromagnetic field in the outer region. Let $\boldsymbol{q}$ be the wave vector of light in the region. The electric field $\boldsymbol{E}'''$ is expressed as

$$\boldsymbol{E}''' = \widehat{\boldsymbol{E}}'''\, e^{i(\boldsymbol{q}\cdot\boldsymbol{r} - \omega t)} \quad (11)$$

where $\widehat{\boldsymbol{E}}'''$ is the constant vector. The electric field satisfies the wave equation. Therefore, $\omega$ and $\boldsymbol{q}$ are related as follows:

$$\boldsymbol{q}\cdot\boldsymbol{q} = \varepsilon_0\mu_0\hat{\varepsilon}_A\hat{\mu}_A\omega^2, \quad (12)$$

The wave vector $\boldsymbol{q}$ is a complex vector owing to the intensity decreasing perpendicular to the travelling direction of the light.

$$\boldsymbol{q} = \boldsymbol{q}^{\text{Re}} + i\boldsymbol{q}^{\text{Im}}, \quad (13)$$

Therein both $\boldsymbol{q}^{\text{Re}}$ and $\boldsymbol{q}^{\text{Im}}$ are vectors with real-number components. Substitution of Eq.(13) into Eq.(12) yields

$$\boldsymbol{q}^{\text{Re}}\cdot\boldsymbol{q}^{\text{Re}} - \boldsymbol{q}^{\text{Im}}\cdot\boldsymbol{q}^{\text{Im}} + 2i\boldsymbol{q}^{\text{Im}}\cdot\boldsymbol{q}^{\text{Re}} = \varepsilon_0\mu_0\hat{\varepsilon}_A\hat{\mu}_A\omega^2 \quad (14)$$

The right-hand side of Eq.(14) is a real-number, because $\hat{\varepsilon}_A \hat{\mu}_A$ for medium A is a real number. Accordingly, vectors $\boldsymbol{q}^{Im}$ and $\boldsymbol{q}^{Re}$ are perpendicular to each other.

$$\boldsymbol{q}^{Im} \cdot \boldsymbol{q}^{Re} = 0 \tag{15}$$

The magnetic field $\boldsymbol{B}'''$ satisfies Maxwell's equations and then the following relation holds:

$$\boldsymbol{B}''' = \frac{1}{\omega} \boldsymbol{q} \times \boldsymbol{E}''' \tag{16}$$

Fig.2  $\boldsymbol{K}$ represents the wave vector inside the prism and $\boldsymbol{q}$ represents the wave vector outside after leaving the prism. Angle $\alpha$ represents the apex angle of the prism. Angle $\delta$ represents the deviation angle through the prism as shown in Fig.1. The two vectors $\boldsymbol{q}^{Re}$ and $\boldsymbol{q}^{Im}$ indicate the real and imaginary parts of $\boldsymbol{q}$. Angles $\chi'$ and $\varphi'$ express the incident and refraction angles at the boundary LHJ.

Figure 2 shows the area near the light exiting the prism. We examined the boundary conditions at the surface LHJ. The parallel component of the electric field to the prism surface is continuous across the boundary. We introduce unit vector $\boldsymbol{u}$ parallel to the line LHJ as shown in Fig. 2. Vector $\boldsymbol{u}$ has the following three components.

$$\boldsymbol{u} = (-\cos\alpha, 0, \sin\alpha) \tag{17}$$

The outside phase $i(\boldsymbol{q}\cdot\boldsymbol{u} - \omega t)$ is the same as the inside phase $i(\boldsymbol{K}\cdot\boldsymbol{u} - \omega t)$ on the boundary LHJ because of the boundary condition of the electric field. Therefore,

$$\boldsymbol{q}\cdot\boldsymbol{u} = \boldsymbol{K}\cdot\boldsymbol{u} \tag{18a}$$

$$q_y = K_y = 0 \tag{18b}$$

The inner products $\boldsymbol{K}^{\text{Re}}\cdot\boldsymbol{u}$ and $\boldsymbol{q}^{\text{Re}}\cdot\boldsymbol{u}$ are expressed by the incident angle $\chi'$ of the light and the refraction angle $\varphi'$ at the boundary as follows:

$$\boldsymbol{K}^{\text{Re}}\cdot\boldsymbol{u} = K^{\text{Re}}\sin\chi' \quad \text{where} \quad K^{\text{Re}} = |\boldsymbol{K}^{\text{Re}}| \tag{19a}$$

$$\boldsymbol{q}^{\text{Re}}\cdot\boldsymbol{u} = q^{\text{Re}}\sin\varphi' \quad \text{where} \quad q^{\text{Re}} = |\boldsymbol{q}^{\text{Re}}| \tag{19b}$$

The vector $\boldsymbol{K}^{\text{Im}}$ has only the z-component. Therefore, Eq.(17) gives

$$\boldsymbol{K}^{\text{Im}}\cdot\boldsymbol{u} = K^{\text{Im}}\sin\alpha \quad \text{where} \quad K^{\text{Im}} = K_z^{\text{Im}} \tag{19c}$$

The vector $\boldsymbol{q}^{\text{Im}}$ is perpendicular to $\boldsymbol{q}^{\text{Re}}$ due to Eq.(15).

Therefore, the following relation is derived from Fig.2:

$$\boldsymbol{q}^{\text{Im}}\cdot\boldsymbol{u} = q^{\text{Im}}\cos\varphi' \quad \text{where} \quad q^{\text{Im}} = |\boldsymbol{q}^{\text{Im}}| \tag{19d}$$

Substitution of these relations into Eq.(18a) yields

$$q^{\text{Re}}\sin\varphi' = K^{\text{Re}}\sin\chi' \tag{20a}$$

$$q^{\text{Im}}\cos\varphi' = K^{\text{Im}}\sin\alpha \tag{20b}$$

In the triangle MGH of Fig. 2, the angle MGH is equal to $\chi$ and the angle MHG is equal to $\chi'$. Therefore, angle $\chi'$ is related to $\alpha$ and $\chi$ as follows:

$$\chi' = \alpha - \chi \tag{21}$$

Substitution of Eq.(15) into (14) gives

$$(q^{Re})^2 - (q^{Im})^2 = \varepsilon_0\mu_0\hat{\varepsilon}_A\hat{\mu}_A\omega^2 \tag{22}$$

(*Short summary 2*): $K^{Re}$ and $K^{Im}$ were determined using Eqs. (8) and (9), and also $\chi'$ is derived from Eqs.(21) and (10b). Therefore, three values $q^{Re}$, $q^{Im}$ and $\varphi'$ can be determined from equations (20a), (20b) and (22) for any value of the incident angle $\varphi$.

Next we calculate the deviation angle $\delta$ shown in Fig.2. Squares of Eqs.(20a) and (20b) yield

$$(q^{Re})^2(\sin\varphi')^2 = (K^{Re})^2(\sin\chi')^2 \tag{23a}$$

$$(q^{Im})^2(\cos\varphi')^2 = (K^{Im})^2(\sin\alpha)^2 \tag{23b}$$

To simplify calculations we introduce new four parameters as follows:

$$X = (q^{Re})^2/(\varepsilon_0\mu_0\hat{\varepsilon}_A\hat{\mu}_A\omega^2) \tag{24a}$$

$$Y = (q^{Im})^2/(\varepsilon_0\mu_0\hat{\varepsilon}_A\hat{\mu}_A\omega^2) \tag{24b}$$

$$f = (K^{Re})^2(\sin\chi')^2/(\varepsilon_0\mu_0\hat{\varepsilon}_A\hat{\mu}_A\omega^2) \tag{24c}$$

$$g = (K^{Im})^2(\sin\alpha)^2/(\varepsilon_0\mu_0\hat{\varepsilon}_A\hat{\mu}_A\omega^2) \tag{24d}$$

Here, the values of $f$ and $g$ have already been determined using Eqs.(9), (10b) and (21). Accordingly our goal is to express the solution of $X$ and $Y$ using the values of $f$ and $g$. Equations (22), (23a) and (23b) yield Eqs.(25a), (25b) and (25c), respectively.

$$X - Y = 1 \rightarrow X = Y + 1 \tag{25a}$$

$$X(\sin\varphi')^2 = f \tag{25b}$$

$$Y(\cos\varphi')^2 = g \tag{25c}$$

Substitution of Eq.(25a) into (25b) yields

$$Y(\sin\varphi')^2 + (\sin\varphi')^2 = f \tag{26}$$

Summation of Eqs.(25c) and (26) yields

$$Y + (\sin\varphi')^2 = f + g$$

$$\rightarrow (\sin\varphi')^2 = f + g - Y \tag{27}$$

Substitution of Eq.(27) into (26) yields

$$Y(f + g - Y) + (f + g - Y) = f$$

$$Y^2 - (f + g - 1)Y - g = 0 \tag{28}$$

Solving the quadratic equation for $Y$

$$Y = \tfrac{1}{2}(f + g - 1) \pm \tfrac{1}{2}\sqrt{(f + g - 1)^2 + 4g}$$

Definition (24d) indicates that the value of $g$ is positive. In addition, the value of $Y$ is positive because of Eq.(25c). Therefore the solution with a negative sign in front of the square root is discarded. Consequently we obtain

$$Y = \tfrac{1}{2}(f + g - 1) + \tfrac{1}{2}\sqrt{(f + g + 1)^2 - 4f} \tag{29a}$$

Substitution of Eq.(29a) into (25a) yields

$$X = \tfrac{1}{2}(f + g + 1) + \tfrac{1}{2}\sqrt{(f + g + 1)^2 - 4f} \tag{29b}$$

Substitution of Eqs.(29b) and (29a) into Eqs.(24a) and (24b) give the values of $q^{\text{Re}}$ and $q^{\text{Im}}$ as;

$$q^{\text{Re}} = \sqrt{\varepsilon_0\mu_0\hat{\varepsilon}_A\hat{\mu}_A\omega^2}\sqrt{\tfrac{1}{2}\left((f + g + 1) + \sqrt{(f + g + 1)^2 - 4f}\right)} \tag{30a}$$

$$q^{\text{Im}} = \sqrt{\varepsilon_0\mu_0\hat{\varepsilon}_A\hat{\mu}_A\omega^2}\sqrt{\tfrac{1}{2}\left((f + g - 1) + \sqrt{(f + g + 1)^2 - 4f}\right)} \tag{30b}$$

Substitution of Eq.(29a) into (27) gives

$$(\sin\varphi')^2 = \tfrac{1}{2}(f + g + 1) - \tfrac{1}{2}\sqrt{(f + g + 1)^2 - 4f} \tag{31}$$

Because the sign of $\sin\varphi'$ is the same as the sign of $\sin\chi'$ owing to Eq.(20a):

$$\sin\varphi' = \sqrt{\tfrac{1}{2}\left(f+g+1-\sqrt{(f+g+1)^2-4f}\right)} \qquad \text{for } \chi' \geq 0 \qquad (32a)$$

$$\sin\varphi' = -\sqrt{\tfrac{1}{2}\left(f+g+1-\sqrt{(f+g+1)^2-4f}\right)} \qquad \text{for } \chi' < 0 \qquad (32b)$$

$$\varphi' = \arcsin\left[\sqrt{\tfrac{1}{2}\left(f+g+1-\sqrt{(f+g+1)^2-4f}\right)}\right] \qquad \text{for } \chi' \geq 0 \qquad (33a)$$

$$\varphi' = -\arcsin\left[\sqrt{\tfrac{1}{2}\left(f+g+1-\sqrt{(f+g+1)^2-4f}\right)}\right] \qquad \text{for } \chi' < 0 \qquad (33b)$$

As shown in Figs.1 and 2, the deviation angle $\delta$ through the prism is

$$\delta = \varphi + \varphi' - \alpha \qquad (34)$$

(*Short summary 3*): In this way, we determined the deviation angle $\delta$ passing through the prism made of a substance with partial light absorption. Equation (33a) and (33b) give the wave vector $\boldsymbol{q}$ via Eqs.(20a), (20b) and (21).

Consequently the electromagnetic fields were determined for all regions on the light pass. The obtained results derive various phenomena that are examined in the next section.

## 3. Minimum deviation angle through Prism

In this section we numerically analyze the refraction phenomena through prisms with a complex number of the dielectric constant. The medium outside the prism is assumed to be a vacuum with $\hat{\varepsilon}_A = 1$, $\hat{\mu}_A = 1$.

The deviation angle $\delta$ through the prism depends on the incident angle $\varphi$. The relationship between $(\varphi, \delta)$ is derived from Eqs.(33a), (33b) and (34). Figure 3 shows four curves of the $(\varphi, \delta)$ relationship. The three lower curves express the $(\varphi, \delta)$ relations for the prisms made of the substance with $\hat{\mu}_B = 1$, $\hat{\varepsilon}_B^{Re} = 3$, $\hat{\varepsilon}_B^{Im} = 0.5$ for three apex angles $\alpha = \pi/4, \pi/5, \pi/6$. The top green curve shows the $(\varphi, \delta)$ relation for the transparent prism with $\hat{\mu}_B = 1$, $\hat{\varepsilon}_B^{Re} = 3$, $\hat{\varepsilon}_B^{Im} = 0$, $\alpha = \pi/4$. This top curve has a critical point and total reflection occurs at $\varphi < 17.0312°$. However, when the prism is made of a substance with $\hat{\varepsilon}_B^{Im} = 0.5$, the total reflection disappears, as shown in the second top curve of Fig.3. Although the second top curve is produced by the prism with the same $\hat{\varepsilon}_B^{Re}$ and same apex angle as in the top curve, the $(\varphi, \delta)$ relation has no critical point. It has a local minimum point and local maximum point as shown in Fig.3. This behavior is unique to prisms that partially absorb light.

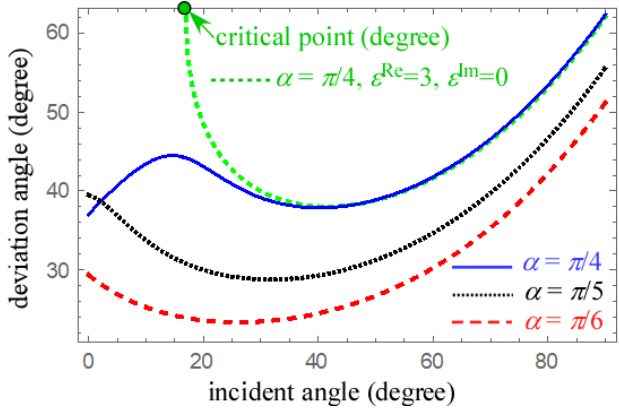

Fig.3 Relation between the deviation angles $\delta$ and the incident angle $\varphi$.

    The top curve indicates $(\varphi, \delta)$ relation for the transparent prism with apex angle $\alpha = \pi/4$. Three lower curves show the deviation angles $\delta$ for the partial absorbing prisms with apex angles $\alpha = \pi/4, \pi/5, \pi/6$. The prisms are made of a substance with $\hat{\mu}_B = 1$ $\hat{\varepsilon}_B^{Re} = 3$, $\hat{\varepsilon}_B^{Im} = 0.5$.

We can numerically determine the local minimum of the deviation angle as shown in Fig.3. In a transparent prism, the refractive index is derived from the minimum deviation angle via Eq.(1). However, Eq.(1) does not provide the refractive index of a prism made of a light-absorbing substance. The minimum deviation angle $\delta_0$ is listed in Table I for $\hat{\varepsilon}_B^{Re} = 3$, $\hat{\varepsilon}_B^{Im} = 0.5$ or $1.0$, $\hat{\mu}_B = 1$ and the apex angles $\alpha = \pi/4, \pi/5, \pi/6, \pi/8$. These eight minimum angles $\delta_0$ are substituted into Eq.(1); the calculated values of $n$ are listed in the fourth column. The results depend on the apex angles as shown in Table I. Thus the traditional method cannot determine the refractive index of media with light absorption.

Table I: List of the values of index $n$ derived from Eq.(1) for eight prisms:

The eight prisms have the apex angles $\alpha = \pi/4, \pi/5, \pi/6, \pi/8$ and are made of substance with $\hat{\varepsilon}_B^{Re} = 3$, $\hat{\varepsilon}_B^{Im} = 0.5$ or $1.0$, $\hat{\mu}_B = 1$. The outside of prism is the vacuum.

| Imaginary Part of a dielectric constant | Apex angle (radian) | Minimum deviation angle (radian) | Index $n$ calculated by Eq.(1) |
|---|---|---|---|
| $\hat{\varepsilon}_B^{Im} = 0.5$ | $\pi/4$ | 0.661132 | 1.72946 |
|  | $\pi/5$ | 0.502498 | 1.73376 |
|  | $\pi/6$ | 0.408031 | 1.73539 |
|  | $\pi/8$ | 0.298606 | 1.73669 |
| $\hat{\varepsilon}_B^{Im} = 1$ | $\pi/4$ | 0.65005 | 1.71858 |
|  | $\pi/5$ | 0.505275 | 1.73755 |
|  | $\pi/6$ | 0.413272 | 1.74443 |
|  | $\pi/8$ | 0.304064 | 1.74984 |

How do we determine the complex refractive index? In the next section we propose a new method for determining the refractive index of a substance with a complex dielectric constant and/or relative permeability.

**4. New Method determining Complex Number of Refraction Index**

The refraction indexes of media A and B are defined by

$$n_A = \sqrt{\hat{\varepsilon}_A \hat{\mu}_A} \tag{35a}$$

$$n_B = \sqrt{(\hat{\varepsilon}_B^{Re} + i\hat{\varepsilon}_B^{Im})(\hat{\mu}_B^{Re} + i\hat{\mu}_B^{Im})} \tag{35b}$$

To simplify calculations, we introduce new parameters $U$ and $V$ as follows:

$$U = (\hat{\varepsilon}_B^{Re} \hat{\mu}_B^{Re} - \hat{\varepsilon}_B^{Im} \hat{\mu}_B^{Im})/(\hat{\varepsilon}_A \hat{\mu}_A) \tag{36a}$$

$$V = (\hat{\varepsilon}_B^{Im} \hat{\mu}_B^{Re} + \hat{\varepsilon}_B^{Re} \hat{\mu}_B^{Im})/(\hat{\varepsilon}_A \hat{\mu}_A) \tag{36b}$$

The quantities $U$ and $V$ depend only on the substances of the prism and the outside medium. The square of $n_B$ is related to $U$ and $V$ as;

$$n_B^2 = n_A^2(U + iV) \tag{37a}$$

$$n_B = n_A\sqrt{U + iV} \tag{37b}$$

New methods are required to determine $U$ and $V$. One method is to measure the two angles in the following two cases:

(Case 1)

To identify Case 1 we attach "case1" to the lower suffices of all quantities. The exit angle $\varphi'_{case1}$ is measured through the prism when the incident angle $\varphi_{case1}$ is zero. Therefore angle $\chi_{case1}$ is also zero owing to Eq.(10b). Accordingly the angle $\chi'_{case1}$ is equal to the apex angle $\alpha$ from Eq.(21).

$$\varphi_{case1} = 0, \ \chi_{case1} = 0 \ \text{and} \ \chi'_{case1} = \alpha \tag{38}$$

In this case, Eq.( 9) becomes

$$\boldsymbol{K}_{case1} = \frac{\omega}{c_0}\left(0, 0, n_A(U^2 + V^2)^{1/4} e^{i\zeta/2}\right) \tag{39a}$$

This equation gives real and imaginary parts as;

$$K^{Re}{}_{case1} = \frac{\omega}{c_0} n_A (U^2 + V^2)^{1/4} \cos(\zeta_{case1}/2) \tag{39b}$$

$$K^{Im}{}_{case1} = \frac{\omega}{c_0} n_A (U^2 + V^2)^{1/4} \sin(\zeta_{case1}/2) \tag{39c}$$

Next we calculate $\cos(\zeta)$ for Case 1 from Eq.(8) as follows;

$$\cos(\zeta_{case1}) = U/\sqrt{U^2 + V^2} \tag{40a}$$

$$\cos^2(\zeta_{case1}/2) = \frac{1}{2}\left(U + \sqrt{U^2 + V^2}\right)/\sqrt{U^2 + V^2} \tag{40b}$$

$$\sin^2(\zeta_{case1}/2) = \frac{1}{2}\left(-U + \sqrt{U^2 + V^2}\right)/\sqrt{U^2 + V^2} \tag{40c}$$

Substituting Eqs.(40b) and (40c) into the squares of Eqs.(39b) and (39c) respectively give the following relations:

$$(K^{Re}{}_{case1})^2 = \left(\frac{\omega}{c_0} n_A\right)^2 \frac{1}{2}\left(U + \sqrt{U^2 + V^2}\right) \tag{41a}$$

$$(K^{Im}{}_{case1})^2 = \left(\frac{\omega}{c_0} n_A\right)^2 \frac{1}{2}\left(-U + \sqrt{U^2 + V^2}\right) \tag{41b}$$

Substitution of Eqs.(41a) and (41b) into Eqs.(24c) and (24d) yields

$$f_{case1} = \frac{1}{2}\left(U + \sqrt{U^2 + V^2}\right)(\sin \alpha)^2 \tag{42a}$$

$$g_{case1} = \frac{1}{2}\left(-U + \sqrt{U^2 + V^2}\right)(\sin \alpha)^2 \tag{42b}$$

$$f_{case1} + g_{case1} = \sqrt{U^2 + V^2}(\sin \alpha)^2 \tag{42c}$$

The measured angle $\varphi'_{case1}$ satisfies the following relation due to Eq.(31).

$$\left(\sin \varphi'_{case1}\right)^2 = \frac{1}{2}\left(\sqrt{U^2 + V^2}(\sin \alpha)^2 + 1\right)$$

$$-\frac{1}{2}\sqrt{(U^2 + V^2)(\sin \alpha)^4 - 2U(\sin \alpha)^2 + 1} \tag{43}$$

The last square root term moves to the left side and all other terms move to the right side. Subsequently, both sides are squared as follows:

$$(U^2 + V^2)(\sin \alpha)^4 - 2U(\sin \alpha)^2 + 1 = \left(\sqrt{U^2 + V^2}(\sin \alpha)^2 + 1 - 2(\sin \varphi'_{case1})^2\right)^2$$

The term $(U^2 + V^2)(\sin \alpha)^4 + 1$ appears on both sides, therefore, simplification of the equation becomes

$$-2U(\sin \alpha)^2 = 2\sqrt{U^2 + V^2}(\sin \alpha)^2(1 - 2(\sin \varphi'_{case1})^2)$$
$$-4(\sin \varphi'_{case1})^2 + 4(\sin \varphi'_{case1})^4$$

Dividing both sides by $(-2(\sin \alpha)^2)$ yields

$$U = -\sqrt{U^2 + V^2}(1 - 2(\sin \varphi'_{case1})^2) + 2((\sin \varphi'_{case1})^2 - (\sin \varphi'_{case1})^4)/(\sin \alpha)^2$$

(44)

Equation (44) expresses the relationship between $U$ and $\sqrt{U^2 + V^2}$ which depends only on the apex angle $\alpha$ and the exit angle $\varphi'_{case1}$. Next we consider another case.

(Case 2)

The second case is specified by $\chi'_{case2} = 0$.

$$\chi'_{case2} = 0 \tag{45a}$$

Equation (24c) gives

$$f_{case2} = 0 \tag{45b}$$

Equation (23a) gives the angle $\varphi'_{case2}$ to be zero:

$$\varphi'_{case2} = 0 \tag{46}$$

Equation (21) yields

$$\chi_{case2} = \alpha \tag{47}$$

The incident angle $\varphi$ is expressed as $\varphi_{case2}$ in Case 2, which is determined by experimental measurement. The value of $\tan \chi_{case2}$ is calculated by using $\varphi_{case2}$ via Eqs.(10a). Substitution of Eq.(47) derives

$$\cot\alpha = \frac{\left((U-\sin^2\varphi_{case2})^2+V^2\right)^{1/4}\cos(\zeta_{case2}/2)}{\sin\varphi_{case2}}$$

which yields

$$\cot^2\alpha = \frac{\left\{(U-\sin^2\varphi_{case2})^2+V^2\right\}^{1/2}(1+\cos(\zeta_{case2}))}{2\sin^2\varphi_{case2}} \tag{48}$$

Equation (8) gives

$$\cos\zeta_{case2} = (U-\sin^2\varphi_{case2})/\sqrt{(U-\sin^2\varphi_{case2})^2+V^2} \tag{49}$$

Substitution of Eq.(49) into (48) yields

$$\cot^2\alpha = \frac{\left\{(U-\sin^2\varphi_{case2})^2+V^2\right\}^{1/2}+(U-\sin^2\varphi_{case2})}{2\sin^2\varphi_{case2}}$$

Multiplying $2\sin^2\varphi_{case2}$ to both sides, we get

$$2\sin^2\varphi_{case2}\cot^2\alpha = \sqrt{(U-\sin^2\varphi_{case2})^2+V^2}+(U-\sin^2\varphi_{case2})$$

Accordingly,

$$\sqrt{(U-\sin^2\varphi_{case2})^2+V^2} = -(U-\sin^2\varphi_{case2})+2\sin^2\varphi_{case2}\cot^2\alpha \tag{50}$$

Square of both sides of Eq.(50) yields

$$V^2 = -4\sin^2\varphi_{case2}\cot^2\alpha\,(U-\sin^2\varphi_{case2})+4\sin^4\varphi_{case2}\cot^4\alpha \tag{51}$$

$U^2$ is added to both sides, then;

$$U^2+V^2 = (-U+2\sin^2\varphi_{case2}\cot^2\alpha)^2+4\sin^4\varphi_{case2}\cot^2\alpha \tag{52}$$

Substitution of Eq.(44) into the right side of Eq.(52) yields

$$U^2+V^2 =$$

$$\left((1-2(\sin\varphi'_{case1})^2)\sqrt{U^2+V^2}+2\sin^2\varphi_{case2}\cot^2\alpha - 2((\sin\varphi'_{case1})^2 - (\sin\varphi'_{case1})^4)/(\sin\alpha)^2\right)^2+4\sin^4\varphi_{case2}\cot^2\alpha \tag{53}$$

To simplify the equation, we introduce parameter $S$ as;

$$S = \sin^2\varphi_{case2} \cot^2\alpha - \left((\sin\varphi'_{case1})^2 - (\sin\varphi'_{case1})^4\right)/(\sin\alpha)^2 \qquad (54)$$

We introduce new coefficients $a, b$ and $c$ as follows:

$$a = 1 - (1 - 2(\sin\varphi'_{case1})^2)^2 \qquad (55a)$$

$$b = -4S \times (1 - 2(\sin\varphi'_{case1})^2) \qquad (55b)$$

$$c = -4S^2 - 4\sin^4\varphi_{case2} \cot^2\alpha \qquad (55c)$$

Equation (53) is represented by using of $a, b$ and $c$ as follows;

$$a\sqrt{U^2+V^2}^2 + b\sqrt{U^2+V^2} + c = 0 \qquad (56)$$

Because of $a > 0$ and $c < 0$, the negative solution of the quadratic equation (56) is discarded. The value of $\sqrt{U^2+V^2}$ becomes;

$$\sqrt{U^2+V^2} = \frac{1}{2a}\left(-b + \sqrt{b^2 - 4ac}\right) \qquad (57)$$

Substitution of Eq.(57) into (44) yields the value of $U$ as

$$U = -\left(1 - 2(\sin\varphi'_{case1})^2\right)\frac{1}{2a}\left(-b + \sqrt{b^2 - 4ac}\right)$$
$$+ 2((\sin\varphi'_{case1})^2 - (\sin\varphi'_{case1})^4)/(\sin\alpha)^2 \qquad (58a)$$

$V$ is derived from Eq.(57) as follows;

$$V = \sqrt{\left(\frac{1}{2a}(-b + \sqrt{b^2 - 4ac})\right)^2 - U^2} \qquad (58b)$$

The measurement of three values $\alpha, \varphi'_{case1}, \varphi_{case2}$ yields $U$ based on Eq.(58a). The determined value of $U$ is substituted on the right-hand side of Eq.(58b), and the value of $V$ is obtained. Consequently, the quantities $(U, V)$ determine the complex refractive index $n_B$ via Eq.(37b).

## 5. Numerical determination of refractive index $n_B$

When medium A is transparent, the complex value of the refractive index $n_B$ has been derived from three measured angles $\alpha, \varphi'_{case1}$ and $\varphi_{case2}$ as studied in the previous section. We numerically calculate the values $\varphi'_{case1}$ and $\varphi_{case2}$ for eight prisms with $\hat{\varepsilon}_B^{Re} = 3, \hat{\varepsilon}_B^{Im} = 0.5$ and 1, and with apex angles $\alpha = \pi/6, \pi/7, \pi/8, \pi/10$. The calculations employed the Mathematica program with a 16-digit system. The obtained values $\varphi'_{case1}$ and $\varphi_{case2}$ are listed on the third and fourth columns of Table II. In the real experiment, these values are obtained from the measurements. The purpose of this study is to find a new method for determining the complex value of the refractive index. Therefore, instead of the measurements, we verify whether the complex refractive index can be numerically derived from the three values in the second, third and fourth columns of Table II.

The values $(U, V)$ were calculated from the values in the third and fourth columns using Eqs.(58a) and (58b). The calculated $(U, V)$ values are presented in the fifth and sixth columns, respectively. The complex index of refraction was evaluated and the results are shown in the seventh column of Table II. The determined index of refraction is independent of the apex angle $\alpha$. The method presented in this paper provides accurate results for determining the complex value of the refractive index.

Table II. Determination of refractive index from measured angles $\varphi'_{case1}$ and $\varphi_{case2}$: The prism is made of a substance with $\hat{\varepsilon}_B^{Re} = 3, \hat{\mu}_B = 1$ and $\hat{\varepsilon}_B^{Im}$. The values of $\hat{\varepsilon}_B^{Im}$ were expressed on the first column. The outside of a prism is the vacuum with $\hat{\varepsilon}_A = 1, \hat{\mu}_A = 1$.

| | $\alpha$ | $\varphi'_{case1}$ (radian) | $\varphi_{case2}$ (radian) | U | V | Refractive index $n_B$ |
|---|---|---|---|---|---|---|
| $\hat{\varepsilon}_B^{Im} = 0.5$ | $\pi/6$ | 1.0362 | 1.05518 | 3 | 0.5 | 1.73801 + $i$ 0.143842 |
| | $\pi/7$ | 0.849201 | 0.855181 | 3 | 0.5 | 1.73801 + $i$ 0.143842 |
| | $\pi/8$ | 0.725242 | 0.728159 | 3 | 0.5 | 1.73801 + $i$ 0.143842 |
| | $\pi/10$ | 0.566086 | 0.567195 | 3 | 0.5 | 1.73801 + $i$ 0.143842 |
| | | | | | | |
| $\hat{\varepsilon}_B^{Im} = 1$ | $\pi/6$ | 1.01154 | 1.07876 | 3 | 1 | 1.75532 + $i$ 0.284849 |
| | $\pi/7$ | 0.846103 | 0.869263 | 3 | 1 | 1.75532 + $i$ 0.284849 |
| | $\pi/8$ | 0.727083 | 0.73853 | 3 | 1 | 1.75532 + $i$ 0.284849 |
| | $\pi/10$ | 0.569823 | 0.57419 | 3 | 1 | 1.75532 + $i$ 0.284849 |

(Note): We numerically calculated all equations using the Mathematica program with a 16-digit system. The values listed in Table II are expressed in eight digits. If the 8 digit values are substituted into the equations, then the calculated results may have some numerical errors.

## 6. Another Method determining Complex Refractive Index

Another method is studied in this section, in which we use two prisms with apex angles $\alpha$ or $\beta$ made of the same substance. The incident angle was chosen to be zero as in Case 1 in Section 4. The third measurement employs a prism with an apex angle $\beta$ that is different from $\alpha$ in Case 1. The measurement is named as Case 3.

**(Case 3)**

$$\varphi_{case3} = 0, \chi_{case3} = 0 \text{ and } \chi'_{case3} = \beta \tag{59}$$

To simplify the calculation, we introduce two parameters $T_{case1}$ and $T_{case3}$ as;

$$T_{case1} = (\sin^2 \varphi'_{case1} - \sin^4 \varphi'_{case1}) / \sin^2 \alpha \tag{60a}$$

$$T_{case3} = (\sin^2 \varphi'_{case3} - \sin^4 \varphi'_{case3}) / \sin^2 \beta \tag{60b}$$

Then Eq.(44) is re-expressed by using Eq.(60a) as

$$U = -\sqrt{U^2 + V^2}(1 - 2(\sin \varphi'_{case1})^2) + 2T_{case1} \tag{61a}$$

The measurement using the prism with apex angle β gives the following relationship similar to Case 1:

$$U = -\sqrt{U^2 + V^2}(1 - 2(\sin \varphi'_{case3})^2) + 2T_{case3} \tag{61b}$$

Subtraction of Eq.(61b) from (61a) for both sides yields

$$0 = 2(\sin^2 \varphi'_{case1} - \sin^2 \varphi'_{case3})\sqrt{U^2 + V^2} + 2(T_{case1} - T_{case3})$$

Then, the solution $\sqrt{U^2 + V^2}$ is expressed as;

$$\sqrt{U^2 + V^2} = (T_{case3} - T_{case1})/(\sin^2 \varphi'_{case1} - \sin^2 \varphi'_{case3}) \tag{62}$$

Summation of both sides of Eqs.(61a) and (61b) yields;

$$U = \sqrt{U^2 + V^2}(-1 + (\sin \varphi'_{case1})^2 + (\sin \varphi'_{case3})^2) + (T_{case1} + T_{case3}) \tag{63}$$

Substitution of Eq.(62) into the right side of Eq.(63) yields;

$$U = \frac{(T_{case3} - T_{case1})(-1 + \sin^2 \varphi'_{case1} + \sin^2 \varphi'_{case3})}{(\sin^2 \varphi'_{case1} - \sin^2 \varphi'_{case3})} + (T_{case1} + T_{case3}) \tag{64}$$

Let $Z$ be the square of right side of Eq.(64) as:

$$Z = \left[\frac{(T_{case3} - T_{case1})(-1 + \sin^2 \varphi'_{case1} + \sin^2 \varphi'_{case3})}{(\sin^2 \varphi'_{case1} - \sin^2 \varphi'_{case3})} + T_{case1} + T_{case3}\right]^2 \tag{65}$$

The value of $V$ is given by

$$V = \sqrt{\frac{(T_{case3} - T_{case1})^2}{(\sin^2 \varphi'_{case1} - \sin^2 \varphi'_{case3})^2} - Z} \tag{66}$$

Consequently, the refractive index of substance B is derived from Eq.(37b) using Eqs.(64) and (66).

Table III. Determination of refractive index from the angles $\varphi'_{case1}$ and $\varphi'_{case3}$: Calculated values on the fourth, fifth and sixth columns are derived from Eqs.(64), (66) and (37b). The substance of the prism has $\hat{\varepsilon}_B^{Re} = 3, \hat{\varepsilon}_B^{Im} = 0.5$, $\hat{\mu}_B = 1$. The outside of prism is the vacuum with $\hat{\varepsilon}_A = 1$, $\hat{\mu}_A = 1$.

| $\alpha$ and $\beta$ radian | $\varphi'_{case1}$ radian | $\varphi'_{case3}$ radian | $U$ from (64) | $V$ from (66) | $n_B$ from (37b) |
|---|---|---|---|---|---|
| $\alpha = \pi/3$ $\beta = \pi/5$ | 1.46103 | 1.31731 | 3 | 0.5 | $1.73801 + i\, 0.143842$ |
| $\alpha = \pi/4$ $\beta = \pi/6$ | 1.43192 | 1.0362 | 3 | 0.5 | $1.73801 + i\, 0.143842$ |
| $\alpha = \pi/5$ $\beta = \pi/8$ | 1.31731 | 0.725242 | 3 | 0.5 | $1.73801 + i\, 0.143842$ |
| $\alpha = \pi/7$ $\beta = \pi/10$ | 0.849201 | 0.566086 | 3 | 0.5 | $1.73801 + i\, 0.143842$ |

The refractive index was numerically calculated using the method examined in this section for several examples. The angles of $\varphi'_{case1}$ and $\varphi'_{case3}$ are listed on the second and third columns in Table III. These two angles determine the values of $(U, V)$ via Eqs.(64) and (66), respectively. The calculated results are listed in the fourth and fifth columns of Table III. The refractive indices are shown in the sixth column. The determined index is independent of the apex angles of prisms and is the same as the result determined in the previous section (please compare with Table II). The method described in this section provides the correct value for the complex index of refraction.

## 7. Conclusions

A new structure appears in the refraction phenomenon when light passes through a prism that partially absorbs it. The optical path through the light absorption prism was determined by solving Maxwell's equations. The exit wave from the prism is different from the normal plane wave as shown in Eqs.(11), (14), (15) and (16). The wave vector outside the prism has an imaginary part $\boldsymbol{q}^{\text{Im}}$ even in a vacuum. The imaginary part $\boldsymbol{q}^{\text{Im}}$ is perpendicular to the real part $\boldsymbol{q}^{\text{Re}}$. The deviation angle $\delta$ of light has a peculiar dependence on the incident angle $\varphi$ for a prism composed of a light-absorbing medium. As shown in Fig.3, the relationship $(\varphi, \delta)$ for a prism made of a substance that partially absorbs light is very different from that for a light-transmitting medium. We found two methods to determine the complex value of the refractive index for a light-absorbing medium. The structure revealed in this study plays an important role in optical systems with partially light-absorbing media.

**Acknowledgements**

The author would like to thank Prof. K. Katsumata for useful discussions.